\documentclass[conference,a4paper]{IEEEtran}
\newif\ifRenderTikZ

\RenderTikZfalse

\ifCLASSINFOpdf
  \usepackage[pdftex]{graphicx}
\else
  \usepackage[dvips]{graphicx}
\fi

\ifCLASSOPTIONcompsoc
 \usepackage[caption=false,font=normalsize,labelfont=sf,textfont=sf]{subfig}
\else
 \usepackage[caption=false,font=footnotesize]{subfig}
\fi

\usepackage{amsfonts}
\usepackage{amsmath}
\usepackage{cite}
\usepackage{xcolor}
\usepackage{balance}
\usepackage{comment}
\usepackage{placeins}
\usepackage{tikz}
\usepackage{pgfplots}
\usepackage{lscape}
\usepackage{xurl}
\usepackage[acronym]{glossaries-extra}
\usepackage{orcidlink}
\usepackage{hyperref}
\hypersetup{
    breaklinks=true,
    pdfauthor={Alexander Ihlow},
    pdftitle={Background Subtraction with Drift Correction for Bistatic Radar Reflectivity Measurements},
    pdfkeywords={Bistatic radar, radar measurements, radar reflectivity, anechoic chambers, network analyzers}
}

\pgfplotsset{compat=newest}

\pgfplotsset{plot coordinates/math parser=false}
\pgfplotsset{footnotesize}
\usepgfplotslibrary{polar,groupplots}

\usetikzlibrary{decorations.shapes, shapes.geometric}
\usetikzlibrary{pgfplots.polar}
\usetikzlibrary{positioning,calc}
\usetikzlibrary{shapes}
\usetikzlibrary{decorations.markings, spy, fit, patterns, arrows}
\usetikzlibrary{intersections}

\ifRenderTikZ
\usetikzlibrary{external}
\tikzset{
    external/system call={
        pdflatex \tikzexternalcheckshellescape
        -halt-on-error -interaction=batchmode
        -jobname "\image" "\texsource"
    }
}

\definecolor{grad1}{RGB}{255,230,180} 
\definecolor{grad2}{RGB}{255,200,140}
\definecolor{grad3}{RGB}{255,170,100}
\definecolor{grad4}{RGB}{250,140,70}
\definecolor{grad5}{RGB}{245,110,50}
\definecolor{grad6}{RGB}{240,85,40}
\definecolor{grad7}{RGB}{235,65,35}
\definecolor{grad8}{RGB}{225,50,30}
\definecolor{grad9}{RGB}{210,40,28}
\definecolor{grad10}{RGB}{195,35,27}
\definecolor{grad11}{RGB}{175,30,25}
\definecolor{grad12}{RGB}{155,28,25}
\definecolor{grad13}{RGB}{135,26,24}
\definecolor{grad14}{RGB}{120,24,23}
\definecolor{grad15}{RGB}{105,22,22}
\definecolor{grad16}{RGB}{90,20,20}
\definecolor{grad17}{RGB}{80,30,25}
\definecolor{grad18}{RGB}{70,40,30}   

\pgfplotscreateplotcyclelist{tol18}{
  {grad18, thick, mark=none},
  {grad17, thick, mark=none},
  {grad16, thick, mark=none},
  {grad15, thick, mark=none},
  {grad14, thick, mark=none},
  {grad13, thick, mark=none},
  {grad12, thick, mark=none},
  {grad11, thick, mark=none},
  {grad10, thick, mark=none},
  {grad9,  thick, mark=none},
  {grad8,  thick, mark=none},
  {grad7,  thick, mark=none},
  {grad6,  thick, mark=none},
  {grad5,  thick, mark=none},
  {grad4,  thick, mark=none},
  {grad3,  thick, mark=none},
  {grad2,  thick, mark=none},
  {grad1,  thick, mark=none},
}
\pgfplotscreateplotcyclelist{tol18_tail3}{
  {grad18, thick, mark=none},
  {grad17, thick, mark=none},
  {grad16, thick, mark=none},
  {grad15, thick, mark=none},
  {grad14, thick, mark=none},
  {grad13, thick, mark=none},
  {grad12, thick, mark=none},
  {grad11, thick, mark=none},
  {grad9,  thick, mark=none},
  {grad8,  thick, mark=none},
  {grad7,  thick, mark=none},
  {grad6,  thick, mark=none},
  {grad5, thick, mark=none},
  {grad4, thick, mark=none},
  {grad3, thick, mark=none},
}
\colorlet{grad19}{grad18!75!black}
\colorlet{grad20}{grad15!75!black}
\colorlet{grad21}{grad12!75!black}
\colorlet{grad22}{grad9!75!black}
\colorlet{grad23}{grad6!75!black}
\colorlet{grad24}{grad3!75!black}
\definecolor{lightorange}{RGB}{255,230,180} 
\definecolor{brownend}{RGB}{70,40,30}

\pgfplotsset{
  colormap={orangeBrown}{
    color(0cm)=(lightorange);
    color(1cm)=(brownend)
  }
}

\pgfplotscreateplotcyclelist{tol24}{
  {color of colormap={0/29},  thick, mark=none},
  {color of colormap={1/29},  thick, mark=none},
  {color of colormap={2/29},  thick, mark=none},
  {color of colormap={3/29},  thick, mark=none},
  {color of colormap={4/29},  thick, mark=none},
  {color of colormap={5/29},  thick, mark=none},
  {color of colormap={6/29},  thick, mark=none},
  {color of colormap={7/29},  thick, mark=none},
  {color of colormap={8/29},  thick, mark=none},
  {color of colormap={9/29},  thick, mark=none},
  {color of colormap={10/29}, thick, mark=none},
  {color of colormap={11/29}, thick, mark=none},
  {color of colormap={12/29}, thick, mark=none},
  {color of colormap={13/29}, thick, mark=none},
  {color of colormap={14/29}, thick, mark=none},
  {color of colormap={15/29}, thick, mark=none},
  {color of colormap={16/29}, thick, mark=none},
  {color of colormap={17/29}, thick, mark=none},
  {color of colormap={18/29}, thick, mark=none},
  {color of colormap={19/29}, thick, mark=none},
  {color of colormap={20/29}, thick, mark=none},
  {color of colormap={21/29}, thick, mark=none},
  {color of colormap={22/29}, thick, mark=none},
  {color of colormap={23/29}, thick, mark=none},
  {color of colormap={24/29}, thick, mark=none},
  {color of colormap={25/29}, thick, mark=none},
  {color of colormap={26/29}, thick, mark=none},
  {color of colormap={27/29}, thick, mark=none},
  {color of colormap={28/29}, thick, mark=none},
  {color of colormap={29/29}, thick, mark=none},
}

\newcommand{\singleplotfile}[2][]{
  \pgfplotstableread[col sep=tab]{#2}\datatable
  \pgfplotstablegetelem{0}{theta_deg}\of{\datatable}\pgfmathparse{\pgfplotsretval}\let\THval\pgfmathresult
  \pgfplotstablegetelem{0}{beta_deg}\of{\datatable}\pgfmathparse{\pgfplotsretval}\let\BEval\pgfmathresult
 \begin{axis}[#1]
    \pgfplotstablegetcolsof{\datatable}
    \pgfmathtruncatemacro{\Nmeas}{\pgfplotsretval-4}
    \pgfplotsinvokeforeach{1,...,\Nmeas}{
      \addplot+[mark=none, thick, forget plot]
        table[col sep=tab, header=true, x index=0, y index=##1]{\datatable};
    }
  \end{axis}
}
\newcommand{\ImpulsePeaksData}{fig/IR_peaks/impulse_peaks.dat}
\definecolor{OIblue}{RGB}{0,114,178}
\definecolor{OIorange}{RGB}{230,159,0}
\definecolor{OIgreen}{RGB}{0,158,115}
\definecolor{OIverm}{RGB}{213,94,0}
\definecolor{OIpurple}{RGB}{204,121,167}
\definecolor{OIskyblue}{RGB}{86,180,233}
\definecolor{OIyellow}{RGB}{240,228,66}
\definecolor{OIblack}{RGB}{0,0,0}
\definecolor{OIgray}{RGB}{153,153,153}
\pgfplotscreateplotcyclelist{oi-list}{
  {OIblue, thick, mark=none},
  {OIorange, thick, mark=none},
  {OIgreen, thick, mark=none},
  {OIverm, thick, mark=none},
  {OIpurple, thick, mark=none},
  {OIskyblue, thick, mark=none},
  {OIyellow,  thick, mark=none},
  {OIblack,   thick, mark=none},
  {OIgray,    thick, mark=none},  
}

\pgfplotscreateplotcyclelist{5step}{
{draw=grad3, mark=none, thin},
{draw=grad7, mark=none, thin},
{draw=grad10, mark=none, thin},
{draw=grad13, mark=none, thin},
{draw=grad16, mark=none, thin},
}

\else
\fi

\newlength\figureheight
\newlength\figurewidth

\setabbreviationstyle[acronym]{long-short}

\glsdisablehyper

\PassOptionsToPackage{hyphens}{url}

\newacronym{adc}{ADC}{analog-to-digital converter}
\newacronym{ai}{AI}{artificial intelligence}
\newacronym{aml}{AML}{antenna measurement laboratory}
\newacronym{api}{API}{application programming interface}
\newacronym[longplural=antennas under test]{aut}{AUT}{antenna under test}
\newacronym{bira}{BIRA}{\textit{Bistatic Radar}}
\newacronym{cad}{CAD}{computer aided design}
\newacronym{cmts}{C-MTS}{Celestia Mechatronic Test Systems}
\newacronym{cpcl}{CPCL}{cooperative passive coherent location}
\newacronym{dac}{DAC}{digital-to-analog converter}
\newacronym{dsi}{DSI}{direct signal interference}
\newacronym{emsl}{EMSL}{European Microwave Signature Laboratory}
\newacronym{ff}{FF}{far field}
\newacronym{gnb}{gNB}{base station}
\newacronym{gnss}{GNSS}{global navigation satellite system}
\newacronym{gpio}{GPIO}{general purpose input/output}
\newacronym{hmt}{RF and Microwave Research Group}{RF and Microwave Research Group}
\newacronym{icas}{ICAS}{integrated communication and sensing}
\newacronym{if}{IF}{intermediate frequency}
\newacronym{iot}{IoT}{internet of things}
\newacronym{isac}{ISAC}{integrated sensing and communication}
\newacronym{lamp}{LAMP}{Laboratory of Target Microwave Properties}
\newacronym{lo}{LO}{local oscillator}
\newacronym{los}{LoS}{line of sight}
\newacronym{mec}{MEC}{mobile edge cloud}
\newacronym{mimo}{MIMO}{multiple-input and multiple-output}
\newacronym{ml}{ML}{machine learning}
\newacronym{ms}{MS}{multi-sensor}
\newacronym{mu}{MU}{multi-user}
\newacronym{mvg}{MVG}{Microwave Vision Group}
\newacronym{ncap}{NCAP}{new car assessment program}
\newacronym{nf}{NF}{near field}
\newacronym{ofdm}{OFDM}{orthogonal frequency-division multiplex}
\newacronym{PCA}{PCA}{Principal Component Analysis}
\newacronym{rcs}{RCS}{radar cross section}
\newacronym{rf}{RF}{radio frequency}
\newacronym{ris}{RIS}{reconfigurable intelligent surfaces}
\newacronym{rx}{Rx}{receiver}
\newacronym{scpi}{SCPI}{Standard Commands for Programmable Instruments}
\newacronym{sdr}{SDR}{software-defined radio}
\newacronym{snr}{SNR}{signal-to-noise ratio}
\newacronym{thimo}{ThIMo}{Thuringian Center of Innovation in Mobility}
\newacronym{tui}{TU Ilmenau}{Technische Universität Ilmenau}
\newacronym{tx}{Tx}{transmitter}
\newacronym[shortplural=UAS]{uas}{UAS}{uncrewed aircraft system}
\newacronym{ue}{UE}{user equipment}
\newacronym{utm}{UTM}{uncrewed aircraft systems traffic management}
\newacronym{vista}{VISTA}{\textit{Virtual Road -- Simulation and Test Area}}
\newacronym{vna}{VNA}{vector network analyzer}
\newacronym{vru}{VRU}{vulnerable road user}

\newcommand{\ee}{\mathrm{e}}

\newcommand{\im}{\mathfrak{Im}}
\newcommand{\jj}{\mathrm{j}}
\newcommand{\phase}{\varepsilon}
\newcommand{\fvec}{\mathbf{f}}
\newcommand{\fgm}{\mathbf{Z}_\text{fg}} % foreground measurement (frequency domain)
\newcommand{\bgm}{\mathbf{z}_\text{bg}} % background measurement (time domain)
\newcommand{\re}{\mathfrak{Re}}

\newcommand{\liftcaption}{\vskip -0.42em}

\begin{document}

\title{Background Subtraction with Drift Correction \\ for Bistatic Radar Reflectivity Measurements}

\author{\IEEEauthorblockN{
Alexander~Ihlow\IEEEauthorrefmark{1}\IEEEauthorrefmark{2}%
\raisebox{.5ex}{\orcidlink{0000-0002-9714-4881}},   % 1st author, 1st affiliations
Marius~Schmidt\IEEEauthorrefmark{1}\IEEEauthorrefmark{2}%
\raisebox{.5ex}{\orcidlink{0009-0004-5961-0994}},	% 2nd author, 2nd affiliations
Carsten~Andrich\IEEEauthorrefmark{1}\IEEEauthorrefmark{2}%
\raisebox{.5ex}{\orcidlink{0000-0002-4795-3517}},    % 3rd author, 3rd affiliations
Reiner~S.~Thomä\IEEEauthorrefmark{1}\IEEEauthorrefmark{2}%
\raisebox{.5ex}{\orcidlink{0000-0002-9254-814X}}      % 4th author, 4th affiliations
}                                   

\IEEEauthorblockA{\IEEEauthorrefmark{1}% 1st affiliations
Institute of Information Technology, Technische Universität Ilmenau, Ilmenau, Germany}

\IEEEauthorblockA{\IEEEauthorrefmark{2}Thuringian Center of Innovation in Mobility, Ilmenau, Germany}

\IEEEauthorblockA{\{alexander.ihlow, marius.schmidt, carsten.andrich, reiner.thomae\}@tu-ilmenau.de}
}

% make the title area
\maketitle

\begin{abstract}
Fundamental research on bistatic radar reflectivity is highly relevant, e.g., to the upcoming mobile communication standard 6G, which includes \gls{isac}.
We introduce a model for correcting instrumentation drift during bistatic radar measurements in anechoic chambers.
Usually, background subtraction is applied with the goal to yield the target reflection signal as best as possible while coherently subtracting all signals which were present in both the foreground and background measurement.
However, even slight incoherences
between the foreground and background measurement process deteriorate the result.
We analyze these effects in real measurements in the frequency range 2--18\,GHz, taken with the \gls{bira} measurement facility at TU Ilmenau.
Applying our proposed drift correction model, we demonstrate up to 40\,dB improvement for
the removal of direct line-of-sight antenna crosstalk
over the state of the art.
\end{abstract}

\vskip0.5\baselineskip

\section{Introduction}

Measuring target reflection signals from radar targets in anechoic chambers requires major investments in the facility itself as well as high effort in planning, conducting and evaluating the experiments.
The target reflection signal is several magnitudes weaker than the illumination signal and often in the order of magnitude of (practically unavoidable) parasitic reflections from the chamber.
Therefore, usually background subtraction is applied to yield the target reflection signal as best as possible.

The measurement process is divided into two steps: target measurement and background measurement, i.e., all bistatic \gls{tx} and \gls{rx} constellations have to be taken twice: with and without the target~\cite{fritsch64_procieee_vfs}.
Several hours may elapse for a measurement, and parameters which were initially presumed as time-invariant may have slightly changed, caused by instrument drift and parasitic effects from cabling due to recurrent bending (for antenna positioning) and temperature changes.
As a consequence, the coherence of the background subtraction is limited.

Wideband measurements enable signal resolution in time-domain.
Depending on the used bandwidth and the propagation delays,
%the direct path leakage signal (also termed \glsfirst{dsi}~\cite{garry07_tgrs_dsi}),
the antenna crosstalk (also termed direct path leakage signal or direct signal interference~\cite{garry07_tgrs_dsi}),
the target reflection signal, as well as parasitic reflections can be distinguished.

While in the monostatic and bistatic case target detection and ranging is possible by reflected signals, the forward scattering case places special challenges, as the signal scattered from a target is received as a modulation on top of the direct path signal~\cite{glaser_taes_rcsforward,gashinova13_eurasip_fwscatter}.

Looking at the state of the art, much work has been done on direct signal interference
and clutter
cancellation for operational radar systems.
A special case are radar reflectivity measurements in anechoic chambers, which we address in this paper.
Here, typical measurement times of several hours and frequently bending of RF cabling due to antenna movements induce drift effects that call for special attention in the data postprocessing.
For such measurements, coherent background subtraction, first practically demonstrated in~\cite{fritsch64_procieee_vfs}, is valued as a convenient method for yielding a purified target signal, eliminating antenna crosstalk and parasitic reflections, as long as coherence between foreground and background measurement is ensured.
As the main goal is to overcome the phase coherence problems that appear with increasing frequency between multiple measurements, suggested solutions range from improving measurement conditions~\cite{Jarvis2022WidebandTargets} to 
various computational methods including image edit- and mean subtraction~\cite{Bati2010AdvancedAlgorithms}, or principal component analysis~\cite{Leye2023BackgroundChamber}.
Typically, coherent background subtraction is used
and also recommended by IEEE Std~1502~\cite[pp.\,31\,ff.]{IEEE_rec_rcs_1502-2020}.

The objective of this paper is to model and correct drift effects, improving the coherence of background subtraction. It is organized as follows:
\autoref{sec:longterm}
introduces a long-term measurement, from which actual drift effects can be quantified.
\autoref{sec:model} derives a correction model and applies it on these measurements.
\autoref{sec:dynamic} shows further results on a multi-position long-term measurement, and \autoref{sec:sphere} on a sphere measurement before \autoref{sec:conc} concludes the paper.

\section{Drift Measurement}\label{sec:longterm}
To analyze the consequences of measurement inaccuracies on background subtraction, focusing on drift, we conducted a long-term measurement of a static scenario in the \glsfirst{bira} measurement facility~\cite{andrich25_arxiv_bira} at TU Ilmenau with the parameters given in \autoref{tab:static_measurement_parameters}.

\begin{table}[h]
\centering
\caption{Parameters of static long-term measurement.}
\label{tab:static_measurement_parameters}
\footnotesize
\renewcommand{\arraystretch}{1.1}
\begin{tabular}{ |l|r| }
\hline
antenna constellation & \gls{tx}: fixed, \gls{rx}: fixed \\ \hline
bistatic angle $\beta$ & antennas facing each other, $\beta=180^\circ$ \\ \hline
radar target & none (direct line-of-sight scenario) \\ \hline
instrument & Keysight N5222B/Option 401 (VNA) \\ \hline
RF connection   & Low-loss coaxial RF cables \\ \hline
frequency range & 2 \ldots 18 GHz \\ \hline
frequency spacing & 10 MHz \\ \hline
measurement runs &  6500\\ \hline
measurement time & 18\,h \\ \hline
\end{tabular}
\end{table}

\begin{figure}[h]
\raggedleft
\ifRenderTikZ
\tikzsetnextfilename{phase_dev_stat}
\begin{tikzpicture}[font=\scriptsize,spy using outlines={lens, magnification=5, size=2cm, connect spies}]
\begin{axis}[
   cycle list name=tol18, 
   width=\columnwidth,height=5cm,
   title=phase deviations,
   every axis title/.style={below right,at={(0,1)},draw=black,fill=white},
   xlabel={Frequency (GHz)},
   ylabel={Phase (°)},
   xmin=0, xmax=20,
   ymin=-0.5, ymax=11,
   ytick={0,2,...,10},
   grid=major,
   legend style={at={(current bounding box.south-|current axis.south)},anchor=north, legend columns=4},
]
\foreach \yindex in {18,...,1}{\addplot+[line join=bevel,mark=none,thin,solid] table[x=f_GHz,y index=\yindex] {fig/Linreg/04_phase_error_deg.dat};}
\foreach \yindex in {16,...,1}{\addplot+[line join=bevel,mark=none,thin,dashed] table[x=f_GHz,y index=\yindex] {fig/Linreg/04_regression_deg_endpoints.dat}; }
\node at (axis cs:20,-0.1) [anchor=south east] {0\,h};
\node at (axis cs:20,1.7) [anchor=south east] {1\,h};
\node at (axis cs:20,3.0) [anchor=south east] {2\,h};
\node at (axis cs:20,3.8) [anchor=south east] {3\,h};
\node at (axis cs:20,4.7) [anchor=south east] {4\,h};
\end{axis}
\end{tikzpicture}
\else
\includegraphics[width=\columnwidth]{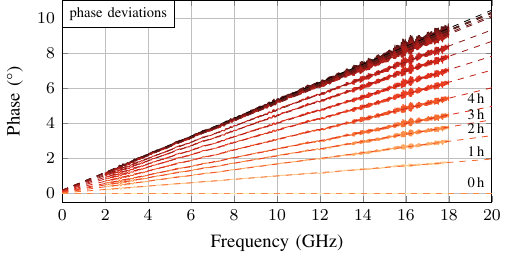}
\fi
\liftcaption
\caption{Phase deviations in static long-term measurement.}
\label{fig:longterm_phase}
\vspace{1em}

\ifRenderTikZ
\tikzsetnextfilename{amplitude_dev_stat}
\begin{tikzpicture}[font=\scriptsize,spy using outlines={lens, magnification=5, size=2cm, connect spies}]
\begin{axis}[
   cycle list name=tol18,
   width=\columnwidth,height=5cm,
   title=amplitude deviations,
   every axis title/.style={below right,at={(0,1)},draw=black,fill=white},
   xlabel={Frequency (GHz)},
   ylabel={Magnitude (dB)},
   xmin=0, xmax=20,
   ymin=-0.01, ymax=0.2,
   ytick={0,0.02,...,0.2},
   yticklabel={\pgfmathprintnumber[fixed, precision=3]{\tick}},
   grid=major,
]
\foreach \yindex in {18,...,2}{\addplot+[line join=bevel,mark=none,thin,solid] table[x=f_GHz,y index=\yindex] {fig/Linreg/03_amplitude_error_dB.dat}; }
\end{axis}
\end{tikzpicture}
\else
\includegraphics[width=\columnwidth]{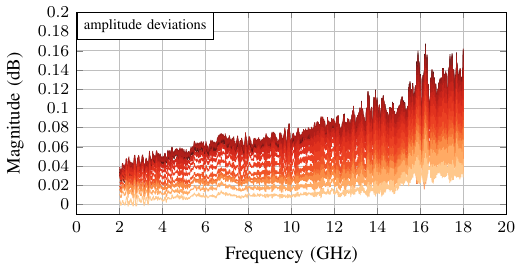}
\fi
\liftcaption
\caption{Amplitude deviations in static long-term measurement.}
\label{fig:longterm_mag}
\end{figure}

\begin{figure}[h]
\centering
\ifRenderTikZ
\tikzsetnextfilename{IR-of-long-term-measurement}
\begin{tikzpicture}[font=\scriptsize,spy using outlines={circle, magnification=5, size=1.5cm, connect spies}]
\begin{axis}[
   cycle list name=tol18,
   width=\columnwidth,height=6.0cm,
   title=impulse response,
   every axis title/.style={below left,at={(1,1)},draw=black,fill=white},
   xlabel={Propagation delay (ns)},
   ylabel={Magnitude (dB)},
   xmin=0, xmax=100,
   xtick={0,10,...,100},
   ymin=-150, ymax=0,
   ytick={-140,-120,...,0},
   grid=major,
]
% Add +10.41 dB offset for normalization to 0 dB
\foreach \yindex in {18,...,2}{\addplot+[line join=bevel,mark=none,thin,solid] table[x expr={\thisrow{sample}/16}, y expr={\thisrowno{\yindex}+10.41}] {fig/Linreg/05_ir_uncorr_dB.dat}; }
\coordinate (spy11) at (axis cs:20,-0);
\coordinate (mag11) at (axis cs:43,-30);
\end{axis}
\spy[gray] on (spy11) in node[fill=white] at (mag11);
\end{tikzpicture}
\else
\includegraphics[width=\columnwidth]{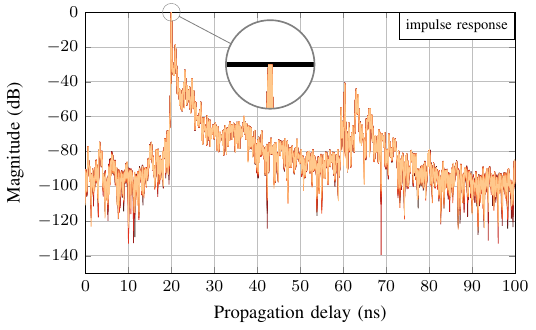}
\fi
\liftcaption
\caption{Impulse responses of static long-term measurement. The set of curves correspond to~\autoref{fig:longterm_phase} and~\autoref{fig:longterm_mag} (cf.~the annotation of measurement time in~\autoref{fig:longterm_phase}).}
\label{fig:longterm_t}
\end{figure}
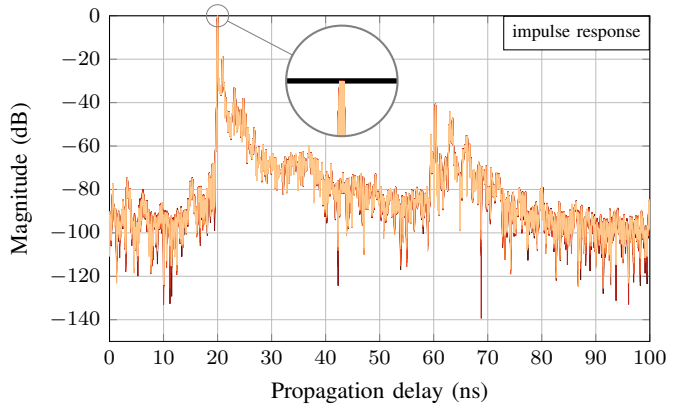

\begin{figure}[h]
\centering
\ifRenderTikZ
\tikzsetnextfilename{IR-bg-subtracted-long-term-measurement}
\begin{tikzpicture}[font=\scriptsize,spy using outlines={lens, magnification=5, size=2cm, connect spies}]
\begin{axis}[
   cycle list name=tol18,
   width=\columnwidth,height=6.0cm,
   title=1\textsuperscript{st} impulse response subtracted from all others,
   every axis title/.style={below left,at={(1,1)},draw=black,fill=white},
   xlabel={Propagation delay (ns)},
   ylabel={Magnitude (dB)},
   xmin=0, xmax=100,
   xtick={0,10,...,100},
   ymin=-150, ymax=0,
   ytick={-140,-120,...,0},
   grid=major,
   legend style={at={(current bounding box.south-|current axis.south)}, anchor=north, legend columns=4},
]
% Add +10.41 dB offset for normalization to 0 dB
\foreach \yindex in {17,...,2}{\addplot+[line join=bevel,mark=none,thin,solid] table[x expr={\thisrow{sample}/16}, y expr={\thisrowno{\yindex}+10.41)}] {fig/Linreg/06_ir_uncorr_bg_subtr_dB.dat}; }
\end{axis}
\end{tikzpicture}
\else
\includegraphics[width=\columnwidth]{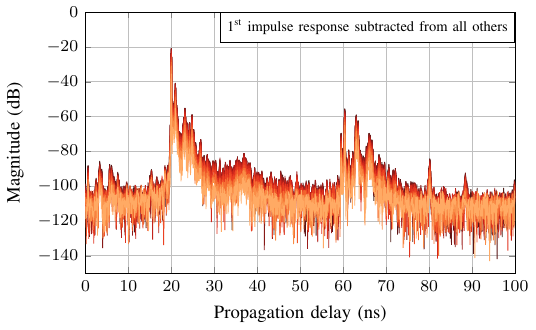}
\fi
\liftcaption
\caption{Impulse responses after conventional background subtraction (\textbf{without} correction): The first measurement is treated as ``background'' and subtracted from all other measurements. The set of curves correspond to~\autoref{fig:longterm_phase} and~\autoref{fig:longterm_mag} (cf.~the annotation of measurement time in~\autoref{fig:longterm_phase}).}
\label{fig:longterm_bg_sub}
\end{figure}
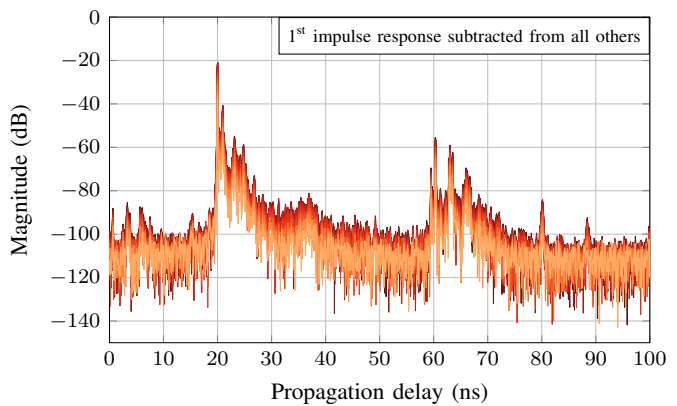

Treating the first measurement as reference, deviations in phase and amplitude are obvious in~\autoref{fig:longterm_phase} and~\autoref{fig:longterm_mag}, increasing with measurement time.
The corresponding impulse responses are drawn in~\autoref{fig:longterm_t}.
To evaluate drift effects on the performance of background subtraction, we use the first measurement as background, while all other measurements are treated as foreground.
The resulting impulse responses after conventional background subtraction are plotted in~\autoref{fig:longterm_bg_sub}, showing a peak residue of $-20$\,dB.
The goal is to develop a model to reduce these residues.

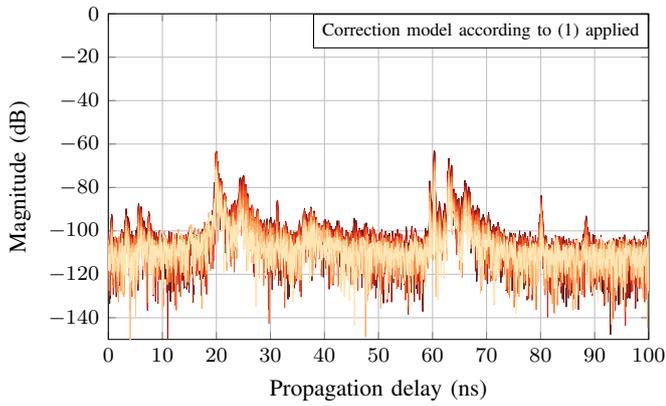
\begin{figure}[h]
\centering
\ifRenderTikZ
\tikzsetnextfilename{Newton-corrected-bg-subtracted-long-term-measurement}
\begin{tikzpicture}[font=\scriptsize,spy using outlines={lens, magnification=5, size=2cm, connect spies}]
\begin{axis}[
   cycle list name=tol18,
   width=\columnwidth,height=6.0cm,
   title=Correction model according to \eqref{eq:model} applied,
   %title = Correction model applied,
   every axis title/.style={below left,at={(1,1)},draw=black,fill=white},
   xlabel={Propagation delay (ns)},
   ylabel={Magnitude (dB)},
   xmin=0, xmax=100,
   xtick={0,10,...,100},
   ymin=-150, ymax=0,
   ytick={-140,-120,...,0},
   grid=major,
   legend style={at={(current bounding box.south-|current axis.south)}, anchor=north, legend columns=4},
]
% Add +10.41 dB offset for normalization to 0 dB
\foreach \yindex in {22,...,4}{\addplot+[line join=bevel,mark=none,thin,solid] table[x expr={\thisrow{sample}/16}, y expr={\thisrowno{\yindex}+10.41)}] {fig/Newton_IR/Newton_Corr_IR.dat}; }
\end{axis}
\end{tikzpicture}
\else
\includegraphics[width=\columnwidth]{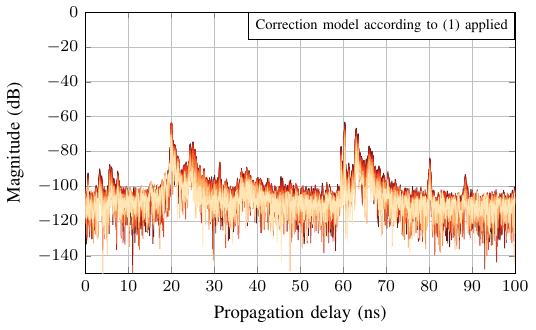}
\fi
\liftcaption
\caption{Impulse responses after background subtraction \textbf{with} correction: The first measurement is treated as ``background'' and subtracted from all other measurements. The model according to \eqref{eq:model} is applied, correcting phase and amplitude deviations. This results in an improvement of up to $40$\,dB regarding peak subtraction.}
\label{fig:longterm_bg_sub_corr}
\end{figure}

\begin{figure}
\raggedleft
\ifRenderTikZ
\tikzsetnextfilename{IR-peak-comparison}
\begin{tikzpicture}[font=\scriptsize,spy using outlines={lens, magnification=5, size=2cm, connect spies}]
  \begin{axis}[
  width=\columnwidth, height=4.5cm,
  xmin=0, xmax = 18,
  ymin=-80, ymax=-5,
  xtick distance=1,
  xticklabel style={
    /pgf/number format/.cd,
      fixed, precision=0,
      1000 sep={},
  },
  yticklabel style={
    /pgf/number format/.cd,
      fixed, precision=0,
  },
  xtick={0,1,...,18},
  ytick={-80,-70,...,-10},
  grid=major,
  title={Impulse response peak value},
  every axis title/.style={below left,at={(1,1)},draw=black,fill=white},
  xlabel={Time (h)},
  ylabel={Magnitude (dB)},
  cycle list name=oi-list,
  legend style={at={(current bounding box.south-|current axis.south)}, anchor=north, legend columns=1},
]
\addplot+[line join=bevel,mark=none,thick] 
    table [col sep=tab, x=time_h, y expr={\thisrow{uncorrected_db}+10.41)}] {\ImpulsePeaksData};
  \addlegendentry{Conventional background subtraction}
\addplot+[line join=bevel,mark=none,thick] 
    table [col sep=tab, x=time_h, y expr={\thisrow{newton_eps_db}+10.41)}] {\ImpulsePeaksData};
    \addlegendentry{Only phase correction (model parameter $\phase$)}
\addplot+[line join=bevel,mark=none,thick] 
    table [col sep=tab, x=time_h, y expr={\thisrow{newton_full_db}+10.41)}] {\ImpulsePeaksData};
    \addlegendentry{Correction model with parameters $\phase, a, b$ applied}

\end{axis}
\end{tikzpicture}
\else
\includegraphics[width=\columnwidth]{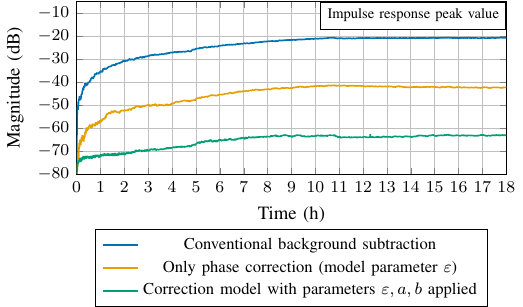}
\fi
\caption{Performance of background subtraction over all static measurement runs (cf.~\autoref{tab:static_measurement_parameters}). The first measurement is treated as ``background''.
The proposed correction model gains up to $40$\,dB peak improvement compared to conventional background subtraction.}
\label{fig:bg_sub_time}
\end{figure}
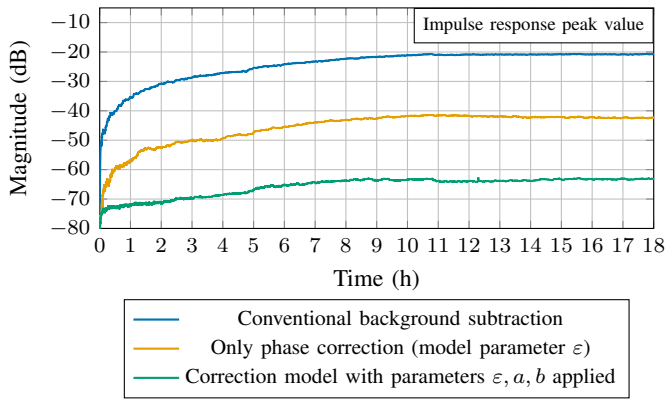

\begin{figure}[h]
\raggedleft
\ifRenderTikZ
\tikzsetnextfilename{Parameter-est-eps}
\begin{tikzpicture}[font=\scriptsize,spy using outlines={lens, magnification=5, size=2cm, connect spies}]
\begin{axis}[
    cycle list name=oi-list, 
   width=\columnwidth,height=3.2cm,
   title=estimated parameter $\phase$ of correction model,
   every axis title/.style={below left,at={(1,1)},draw=black,fill=white},
   ylabel={$\phase$ (°/GHz)},
   xmin=0, xmax = 18,
   xtick distance=1,
    yticklabel={\pgfmathprintnumber[fixed, precision=6]{\tick}},
   scaled x ticks=false,
   ymin=0, ymax=1.0,
   ytick={0,0.2,...,1.0},
   grid=major,
   legend style={at={(current bounding box.south-|current axis.south)},anchor=north, legend columns=2},
]
    \addplot+[line join=bevel,mark=none,thick]table [col sep=tab, x=time_h, y=newton_full_db]{fig/IR_peaks/impulse_peaks_eps.dat};
\end{axis}
\end{tikzpicture}
\else
\includegraphics[width=\columnwidth]{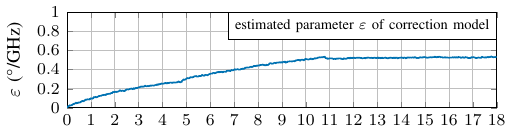}
\fi

\ifRenderTikZ
\tikzsetnextfilename{Parameter-est-a}
\begin{tikzpicture}[font=\scriptsize,spy using outlines={lens, magnification=5, size=2cm, connect spies}]
\begin{axis}[
    cycle list name=oi-list, 
   width=\columnwidth,height=3.2cm,
   title=estimated parameter $a$ of correction model,
   every axis title/.style={below left,at={(1,1)},draw=black,fill=white},
   ylabel={$a$},
   xmin=0, xmax = 18,
   xtick distance=1,
   scaled x ticks=false,
   ymin=0.98, ymax=1.01,
   ytick={0.98,0.99,...,1.01},
   grid=major,
   legend style={at={(current bounding box.south-|current axis.south)},anchor=north, legend columns=2},
]
    \addplot+[line join=bevel,mark=none,thick]table [col sep=tab, x=time_h, y=newton_full_db]{fig/IR_peaks/impulse_peaks_a.dat};
\end{axis}
\end{tikzpicture}
\else
\includegraphics[width=\columnwidth]{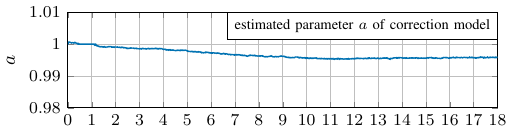}
\fi

\ifRenderTikZ
\tikzsetnextfilename{Parameter-est-b}
\begin{tikzpicture}[font=\scriptsize,spy using outlines={lens, magnification=5, size=2cm, connect spies}]
\begin{axis}[
    cycle list name=oi-list, 
   width=\columnwidth,height=3.2cm,
   title=estimated parameter $b$ of correction model,
   every axis title/.style={below left,at={(1,1)},draw=black,fill=white},
   xlabel={Time (h)},
   ylabel={$b$ (1/GHz)},
   xmin=0, xmax = 18,
   xtick distance=1,
   scaled x ticks=false,
   ymin=-0.0025, ymax=0.0025,
   ytick={-0.0025,-0.0015,...,0.0025},
   grid=major,
      yticklabel={\pgfmathprintnumber[fixed, precision=3]{\tick}},
   legend style={at={(current bounding box.south-|current axis.south)},anchor=north, legend columns=2},
]
    \addplot+[line join=bevel,mark=none,thick]table [col sep=tab, x=time_h, y=newton_full_db]{fig/IR_peaks/impulse_peaks_b.dat};
\end{axis}
\end{tikzpicture}
\else
\includegraphics[width=\columnwidth]{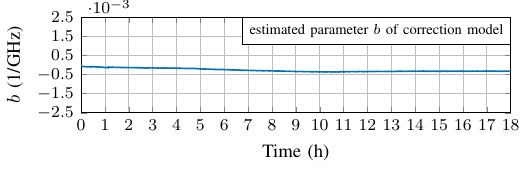}
\fi
\liftcaption
\caption{Estimated parameters for static measurement (cf.~\autoref{tab:static_measurement_parameters}).}
\label{fig:longterm_parameters}
\end{figure}
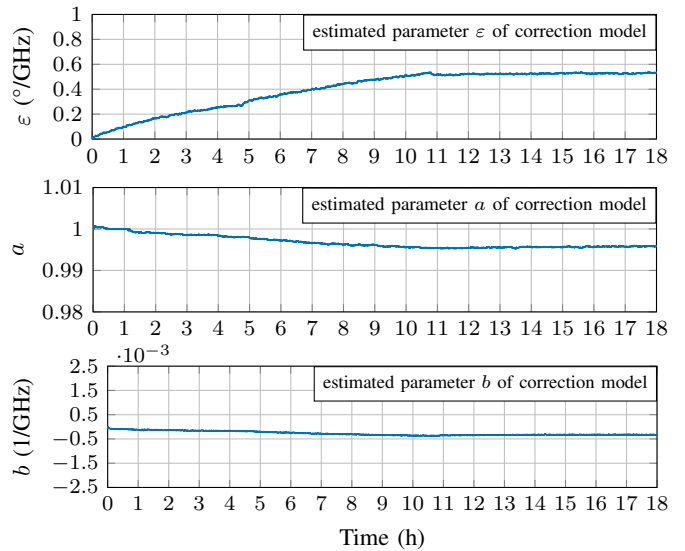

\section{Correction Model}\label{sec:model}
From \autoref{fig:longterm_phase} and \autoref{fig:longterm_mag} it is obvious that the correction model should consider phase as well as amplitude deviations.
Furthermore, it should be able to focus on particular samples of the impulse response, such as the peak from the direct signal (cf.~\autoref{fig:longterm_t}).
We formulate the model as follows:
\begin{align}
    \underset{a,b,\phase \; \in \, \mathbb{R}}{\arg\min} \sum\limits_{n}\left\lvert\text{IDFT}\{(a+b\,\fvec) \, \ee^{-\jj\phase \fvec} \, \fgm\}[n] - \bgm[n]\right\rvert^2 = \label{eq:model} \\
    \underset{a,b,\phase \; \in \, \mathbb{R}}{\arg\min} \sum\limits_{n}\left\lvert\frac{1}{N}\sum\limits_{k=0}^{N-1}(a+b\, f_k) \, \ee^{-\jj\phase f_k} \, \fgm[k] \, \ee^{\jj2\pi\frac{k}{N}n} - \bgm[n]\right\rvert^2 \nonumber
\end{align}
$\fgm$ is the spectrum of a single foreground measurement, $\bgm=\text{IDFT}\{\mathbf{Z}_\text{bg}\}$ is the corresponding (same illumination and observation angles
and polarization) background measurement in time-domain after inverse discrete Fourier transform (IDFT). The operation $[n]$ selects the relevant samples in time domain, on which the parameter optimization is performed.
Of course, all vector multiplications within the $\text{IDFT}\{\cdot\}$ in \eqref{eq:model} are \textit{element-wise} products.
The commonly used symbol $\odot$ is omitted throughout the paper for better readability.

The frequencies $f_k$ can be defined, e.g., in complex baseband or as physical frequencies.
Without loss of generality, we chose
$f_k = f_\text{start} + (f_\text{end}-f_\text{start}) \frac{k}{N}\left(1+\frac{1}{N}\right)$.
In our measurements, $f_\text{start}=2$\,GHz, $f_\text{end}=18$\,GHz, and $N=1601$.
Doing so, the parameter $\phase$ gets the unit °/GHz, $b$ gets the unit 1/GHz, $a$ is dimensionless.
Note that the choice to apply the correction on the foreground measurement is arbitrary, the model would work also vice versa.

For this model, the partial derivatives can be analytically derived (given in the Appendix), making it easily applicable for numerical solution via, e.g., Newton-Conjugate-Gradient.

\newpage
Applying the correction model onto the longterm measurement, the peak residue reduces from $-20$\,dB to $-60$\,dB, as seen in comparison of~\autoref{fig:longterm_bg_sub} and~\autoref{fig:longterm_bg_sub_corr}.
Details on peak reduction over time are drawn in~\autoref{fig:bg_sub_time}, estimated parameters are given in~\autoref{fig:longterm_parameters}.

\section{Multi-Position Measurement}\label{sec:dynamic}
To confirm the effectiveness of this algorithm for more realistic scenarios, 
we conducted a further measurement with parameters according to~\autoref{tab:dynamic_measurement_parameters}.
\begin{table}[h]
\centering
\caption{Parameters of multi-position long-term measurement.}
\label{tab:dynamic_measurement_parameters}
\footnotesize
\renewcommand{\arraystretch}{1.1}
\begin{tabular}{ |l|r| }
\hline
RF parameters & same as in~\autoref{tab:static_measurement_parameters}\\ \hline
antenna constellation & \gls{tx}: fixed, \gls{rx}: variable elevation \\ \hline
bistatic angles $\beta$ & 15 constellations between 30° and 150° \\ \hline
radar target & none (direct line-of-sight scenario) \\ \hline
measurement runs &  288 for each bistatic angle\\ \hline
measurement time & 20\,h \\ \hline
\end{tabular}
\end{table}

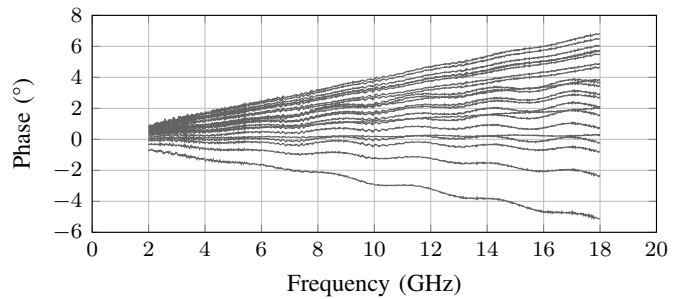
\begin{figure}[h!]
\raggedleft
\ifRenderTikZ
\tikzsetnextfilename{Phase-Dev-multi-pos}
\begin{tikzpicture}[font=\scriptsize]
\begin{axis}[
  width=\columnwidth, height=4.4cm,
  xlabel={Frequency (GHz)}, ylabel={Phase (°)},
  xmin=0, xmax=20,
  ymin=-6, ymax=8,
  enlarge y limits=false,
  ytick={-6,-4,-2,0,2,4,6,8},
  grid=major,
  legend style={at={(0.98,0.98)},anchor=north east},
]

\foreach \h in {22,21,20,19,18,17,16,15,14,13,12,11,10,09,08,07,06,05,04,03,02,01}{
\addplot[mark=none, thin, draw={rgb,255:red,99; green,99; blue,99}]
table[x=f_GHz, y=Phase_Diff_\h h_DEG, col sep=tab]{fig/02_measure_els/Overlay___PhaseDiff_theta90_beta150p00_02_measure_els.dat};
}
\end{axis}
\end{tikzpicture}
\else
\includegraphics[width=\columnwidth]{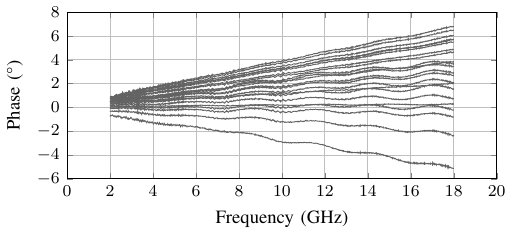}
\fi
\liftcaption
\caption{Phase deviations in multi-position long-term measurement. A clear dependency of the deviations on measurement time, as in~\autoref{fig:longterm_phase} cannot be deduced.
However, harmonic phase modulations appear, which are induced by residual mechanical oscillations of the gantry arm antenna positioner.}
  \label{fig:phase_dev_els}
\end{figure}

\begin{figure}[h!]
\raggedleft
\ifRenderTikZ
\tikzsetnextfilename{Mag-Dev-multi-pos}
\begin{tikzpicture}[font=\scriptsize]
\begin{axis}[
width=\columnwidth,
xlabel={Frequency (GHz)}, ylabel={Magnitude (dB)},
height = 4.4cm,
grid=major, legend style={at={(0.98,0.98)},anchor=north east},xmin=0, xmax=20, ymin=-0.06, ymax=0.12, ytick={-0.06,-0.04,...,0.12}, yticklabel style={/pgf/number format/.cd, fixed, precision=2,
},clip=false,
]

\foreach \h in {22,21,20,19,18,17,16,15,14,13,12,11,10,09,08,07,06,05,04,03,02,01}{
\addplot[mark=none, thin, draw={rgb,255:red,99; green,99; blue,99}]
table[x=f_GHz, y=Magnitude_Diff_\h h_dB, col sep=tab]{fig/02_measure_els/Overlay___MagnitudeDiff_theta90_beta150p00_02_measure_els.dat};
}
\end{axis}
\end{tikzpicture}
\else
\includegraphics[width=\columnwidth]{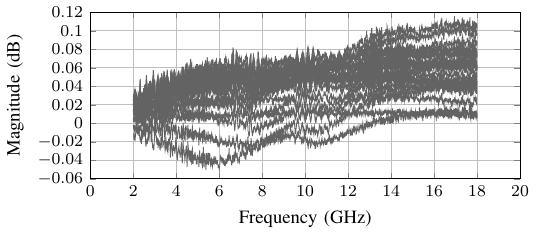}
\fi
\liftcaption
  \caption{Amplitude deviations in multi-position long-term measurement.}

\label{fig:mag_diff_els}
\end{figure}
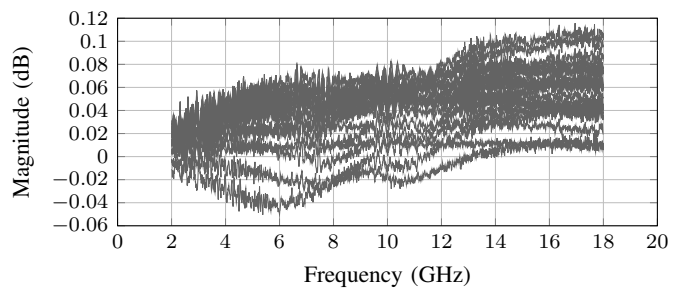

As these data are too comprehensive to be fully evaluated in this paper, we focus on one representative constellation with the bistatic angle $\beta=150^\circ$.
\autoref{fig:phase_dev_els} and~\autoref{fig:mag_diff_els} show further effects not found in the previously analyzed static data, where neither cable bending nor possibly tiny oscillation residues from the gantry arm antenna positioner played a role.
Nevertheless, on these data the correction model provides up to 37\,dB peak improvement in background subtraction.

\newpage
\section{Application onto Sphere Measurements}\label{sec:sphere}
We applied the proposed algorithm onto measurements of an azimuth cut of a 30\,cm sphere in horizontal co-polarization (described in~\cite{andrich25_arxiv_bira} and summarized in~\autoref{tab:sphere_measurement_parameters}).

\begin{table}[h]
\centering
\caption{Parameters of sphere measurement.}
\label{tab:sphere_measurement_parameters}
\footnotesize
\renewcommand{\arraystretch}{1.1}
\begin{tabular}{ |l|r| }
\hline
RF parameters & same as in~\autoref{tab:static_measurement_parameters}\\ \hline
antenna constellation & \gls{tx}: variable azimuth, \gls{rx}: fixed \\ \hline
bistatic angles $\beta$ & $\beta = 10^\circ \ldots 243^\circ$ in $1^\circ$ azimuth steps \\ \hline
radar target & 30\,cm sphere \\ \hline
measurement time & 1\,h 20\,min foreground + 1\,h 20\,min background \\ \hline
\end{tabular}
\end{table}

As stated in \autoref{sec:model}, each corresponding pair (same illumination and observation angles as well as polarization) of foreground and background measurement is corrected according to~\eqref{eq:model}.
The selected samples $n$ in time domain for which the parameter optimization is performed, are the respective direct signal peak samples within the impulse response of the background measurement, found by $\underset{n\,\in\,0\ldots N-1}{\arg\max}\{|\bgm[n]|\}$.

\begin{figure*}
    \centering
    
\begin{tikzpicture}
\pgfdeclarelayer{foreground}
\pgfsetlayers{main,foreground}
\begin{axis}[
        domain=10:243,
        colorbar,
        colormap/jet,
        colorbar style={
            ylabel={Reflectivity (dBsm)},
            ytick={-80,-60,...,10},
            yticklabel style={
                text width=1.5em,
                align=right
            }
        },
        width=0.89\linewidth,
        height=4.4cm,
        enlargelimits=false,
        axis on top,
        tick align=outside,
        tick pos=left,
        xtick pos=top,
        title={Conventional background subtraction},
        every axis title/.style={above right,at={(0,0)},draw=black,fill=white},
        ylabel={Path length (m)},
        xtick distance=15,
        ytick distance=0.2,
        x tick label style={rotate=90,anchor=west}
    ]
    \addplot[point meta min=-80, point meta max=20] graphics[xmin=10, xmax=243, ymin=-0.9375, ymax=0.5625] {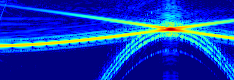};
    
\end{axis}
\end{tikzpicture}

\begin{tikzpicture}
\pgfdeclarelayer{foreground}
\pgfsetlayers{main,foreground}
\begin{axis}[
        domain=10:243,
        colorbar,
        colormap/jet,
        colorbar style={
            ylabel={Reflectivity (dBsm)},
            ytick={-80,-60,...,10},
            yticklabel style={
                text width=1.5em,
                align=right
            }
        },
        width=0.89\linewidth,
        height=4.4cm,
        enlargelimits=false,
        axis on top,
        tick align=outside,
        tick pos=left,
        xtick pos=bottom,
        title={Proposed correction model applied},
        every axis title/.style={above right,at={(0,0)},draw=black,fill=white},
        xlabel={Bistatic azimuth angle (°)},
        ylabel={Path length (m)},
        xtick distance=15,
        ytick distance=0.2,
        x tick label style={rotate=90,anchor=east}
    ]
    \addplot[point meta min=-80, point meta max=20] graphics[xmin=10, xmax=243, ymin=-0.9375, ymax=0.5625] {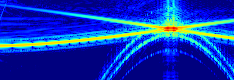};
\end{axis}
\end{tikzpicture}
    \caption{
        Bistatic azimuth cut of a 30\,cm sphere, measured in the \gls{bira} facility from 2 to 18\,GHz,  horizontal to horizontal (HH) polarization.
        Top plot with conventional background subtraction (foreground -- background), bottom plot with the proposed correction model according to~\eqref{eq:model} applied.
    }
    \label{fig:meas_hh}
\end{figure*}

\begin{figure}
\raggedleft
\ifRenderTikZ
\tikzsetnextfilename{Sphere-epsilon}
\begin{tikzpicture}[font=\scriptsize,spy using outlines={lens, magnification=5, size=1.2cm, connect spies}]
\begin{axis}[
   width=\columnwidth,height=3.2cm,
   title=parameter $\phase$,
   every axis title/.style={below right,at={(0.01,0.7)},draw=black,fill=white},
   ylabel={$\phase$ (°/GHz)},
   xmin=10, xmax=243,
   ymin=-10, ymax=2,
   ytick={-10,-8,...,2},
   grid=major,
]
\addplot+[line join=bevel,mark=none,thick,blue,solid] table[x=ind,y=hh] {fig/sphere/eps.dat};
\coordinate (spy11) at (axis cs:160,0);
\coordinate (mag11) at (axis cs:100,-5);
\end{axis}
\end{tikzpicture}
\else
\includegraphics[width=\columnwidth]{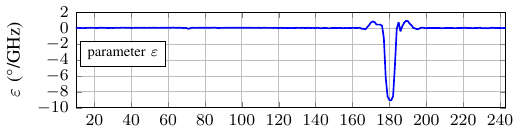}
\fi

\ifRenderTikZ
\tikzsetnextfilename{Sphere-a}

\begin{tikzpicture}[font=\scriptsize]
\begin{axis}[
   width=\columnwidth,height=3.2cm,
   title=parameter $a$,
   every axis title/.style={below right,at={(0.01,0.7)},draw=black,fill=white},
   ylabel={$a$},
   xmin=10, xmax=243,
   ymin=0.99, ymax=1.03,
   ytick={0.99,1.0,...,1.03},
   grid=major,
]
\addplot+[line join=bevel,mark=none,thick,blue,solid] table[x=ind,y=hh] {fig/sphere/a.dat};
\end{axis}
\end{tikzpicture}
\else
\includegraphics[width=\columnwidth]{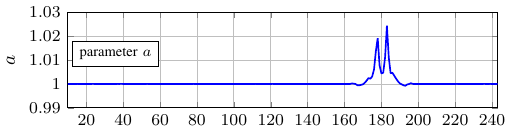}
\fi

\ifRenderTikZ
\tikzsetnextfilename{Sphere-b}
\begin{tikzpicture}[font=\scriptsize]
\begin{axis}[
   width=\columnwidth,height=3.2cm,
   title=parameter $b$,
   every axis title/.style={below right,at={(0.01,0.7)},draw=black,fill=white},
   xlabel={Bistatic angle $\beta$ (°)},
   ylabel={$b$ (1/GHz)},
   xmin=10, xmax=243,
   ymin=-0.1, ymax=0.3,
   ytick={-0.1,0,...,0.3},
   grid=major,
]
\addplot+[line join=bevel,mark=none,thick,blue,solid] table[x=ind,y=hh] {fig/sphere/b.dat};
\end{axis}
\end{tikzpicture}
\else
\includegraphics[width=\columnwidth]{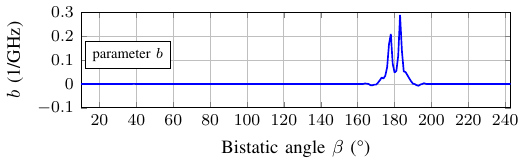}
\fi
\liftcaption
\caption{Estimated model parameters on sphere measurements. \newline In the pseudo-monostatic and bistatic region, the model shows valid results. \newline In the forward scattering region ($\beta\approx160-200$°), the model is not applicable
as the direct line-of-sight
antenna crosstalk is no longer distinguishable from target interaction signals in the impulse responses.
Here, the model parameters deviate far from plausible values, cf.~\autoref{fig:longterm_parameters}.}
\label{fig:sphere_parameters}
\end{figure}

Results are drawn in~\autoref{fig:meas_hh} and~\autoref{fig:sphere_parameters}.
The algorithm works well in the bistatic region, removing the residue of the direct path leakage signal.
However, in the forward scattering region ($\beta \approx 160^\circ \ldots 200^\circ$), where the direct line-of-sight antenna crosstalk
can no longer be distinguished from target signals, it is not applicable.
Here, the model parameters deviate far from plausible values (cf.~\autoref{fig:sphere_parameters}), distorting the correctness of estimated bistatic radar reflectivity values.

\section{Conclusion}\label{sec:conc}
We investigated instrument drift in bistatic radar reflectivity measurements in the \gls{bira} measurement facility~\cite{andrich25_arxiv_bira} at TU Ilmenau.
As expected, technically unavoidable deviations in phase and amplitude limit the coherence between foreground and background measurement, even with cutting-edge hardware.
This manifests mainly as a residue of the direct signal (line-of-sight antenna crosstalk) after background subtraction.
To improve the coherence, we introduced a parametric correction model, containing a phase term and an amplitude term.
Model parameter estimation is performed for all corresponding pairs of foreground and background measurements by focusing on the direct line-of-sight signal portion of the impulse responses.
Newton-Conjugate-Gradient ensures fast convergence, using the analytically expressed partial derivatives.
Regarding direct signal suppression, our algorithm yields up to 40\,dB improvement compared to conventional background subtraction.
The model is applicable in angular constellations where the direct antenna crosstalk can be distinguished from all target interaction signals in time domain, which is the monostatic and bistatic region.
In the forward scattering region, where this demand is not met, conventional background subtraction can be applied.

Whether an interpolation of estimated parameters from the monostatic and bistatic region could be relied upon in the forward scattering region is subject of further investigation.

%\appendices %IEEEtran special command!
\section*{Appendix} \label{app}

To find the minimum of the function
\begin{align}
    \sum\limits_{n}\left\lvert\text{IDFT}\{(a+b\, \fvec) \, \ee^{-\jj\phase \fvec} \, \fgm\}[n] - \bgm[n]\right\rvert^2 \nonumber  = \sum\limits_{n}| \kappa \rvert^2 \\
=
\sum\limits_{n}\left\lvert\frac{1}{N}\sum\limits_{k=0}^{N-1}(a+b\,f_k) \, \mathrm{e}^{-\mathrm{j}\phase f_k} \, \fgm[k] \, \ee^{\jj2\pi\frac{k}{N}n} - \bgm[n]\right\rvert^2
\end{align}
with respect to the parameters $\phase, a, b \in \mathbb{R}$, we use $|\kappa|^2 = \re\{\kappa\}^2 + \im\{\kappa\}^2$ with $\kappa \in \mathbb{C}$. To express the first and second derivatives in general, we use the chain rule and $p, \nu \in \mathbb{R}$:

\begin{align}
\label{eq:first_deriv}
\frac{\partial |\kappa|^2}{\partial p}
&= 2\,\re\{\kappa\}\,\frac{\partial}{\partial p}\re\{\kappa\} + 2\,\im\{\kappa\}\,\frac{\partial}{\partial p}\im\{\kappa\} %\\
\end{align}

\begin{align}
\label{eq:second_deriv}
\frac{\partial^2 |\kappa|^2}{\partial p \, \partial \nu} 
&= 2 \, \frac{\partial}{\partial \nu}\re\{\kappa\} \, \frac{\partial}{\partial p}\re\{\kappa\} 
   + 2 \, \re\{\kappa\} \, \frac{\partial^2}{\partial p \, \partial \nu}\re\{\kappa\} \notag \\
&\,+ 2 \, \frac{\partial}{\partial \nu}\im\{\kappa\} \, \frac{\partial}{\partial p}\im\{\kappa\} 
   + 2 \, \im\{\kappa\} \, \frac{\partial^2}{\partial p \, \partial \nu}\im\{\kappa\}
\end{align}
Explicitly, the first and second derivatives in \eqref{eq:first_deriv} and \eqref{eq:second_deriv} are:
\begin{align}
    \frac{\partial}{\partial \phase} \Big[\kappa \Big]
&= \frac{1}{N}\sum\limits_{k=0}^{N-1}(a+b\,f_k)\,(-\jj\,f_k) \,\mathrm{e}^{-\mathrm{j}\phase f_k} \, \fgm[k] \, \ee^{\jj2\pi\frac{k}{N}n} \nonumber \\
&= \text{IDFT}\{(a+b\,\fvec)\,(-\jj\,\fvec) \,\mathrm{e}^{-\mathrm{j}\phase \fvec} \, \fgm\} \\
    \frac{\partial}{\partial a} \Big[\kappa \Big]
&= \frac{1}{N}\sum\limits_{k=0}^{N-1}\ee^{-\jj\phase f_k} \, \fgm[k] \, \ee^{\jj2\pi\frac{k}{N}n} \nonumber \\
&= \text{IDFT}\{\ee^{-\jj\phase \fvec} \, \fgm\} \\
    \frac{\partial}{\partial b} \Big[\kappa \Big]
&= \frac{1}{N}\sum\limits_{k=0}^{N-1}f_k\,\ee^{-\jj\phase f_k} \, \fgm[k] \, \ee^{\jj2\pi\frac{k}{N}n} \nonumber \\
&= \text{IDFT}\{\fvec\,\ee^{-\jj\phase \fvec} \, \fgm\}
\end{align}
\begin{align}
\frac{\partial^2}{\partial \phase^2}\big[\kappa\big]
&= \frac{1}{N}\sum_{k=0}^{N-1} (a+b f_k)(-f_k^2)\,\mathrm{e}^{-\jj \phase f_k}\,\fgm[k]\,\mathrm{e}^{\jj 2\pi \tfrac{k}{N} n} \notag \\
&= \mathrm{IDFT}\!\left\{(a+b\,\fvec)(-\fvec^2)\,\mathrm{e}^{-\jj \phase \fvec}\,\fgm\right\} \\
\frac{\partial^2}{\partial a \, \partial \phase}\big[\kappa\big]
&= \frac{1}{N}\sum_{k=0}^{N-1} \big(-\jj f_k\big)\,\mathrm{e}^{-\jj \phase f_k}\,\fgm[k]\,\mathrm{e}^{\jj 2\pi \tfrac{k}{N} n} \notag \\
&= \mathrm{IDFT}\!\left\{-\jj \fvec\,\mathrm{e}^{-\jj \phase \fvec}\,\fgm\right\} \\
\frac{\partial^2}{\partial \phase \, \partial b}\big[\kappa\big]
&= \frac{1}{N}\sum_{k=0}^{N-1} \big(-\jj f_k^2\big)\,\mathrm{e}^{-\jj \phase f_k}\,\fgm[k]\,\mathrm{e}^{\jj 2\pi \tfrac{k}{N} n} \notag \\
&= \mathrm{IDFT}\!\left\{-\jj \fvec^2\,\mathrm{e}^{-\jj \phase \fvec}\,\fgm\right\} \\
\frac{\partial^2}{\partial a^2}\big[\kappa\big] &= 
\frac{\partial^2}{\partial b^2}\big[\kappa\big] =  \frac{\partial^2}{\partial b \, \partial a}\big[\kappa\big] = \frac{\partial^2}{\partial a \, \partial b}\big[\kappa\big] = 0
\end{align}
With these expressions it is straightforward to find the minimum, e.g., with the Newton-Conjugate-Gradient algorithm.

\section*{Reference Implementation of the Algorithm}
A reference implementation in the Python programming language of the core function containing the proposed model and optimization is published on \href{https://arxiv.org/}{arxiv.org}.

\section*{Acknowledgment}
The research was funded by the Federal State of Thuringia, Germany, and the European Social Fund (ESF) under grants 2017 FGI 0007 (project ``BiRa''), 2021 FGI 0007 (project ``Kreatör''), and 2023 IZN 0005 (project ``research initiative digital mobility''). 
We thank Dr. Tobias Nowack for supporting the measurements and
Lisa Hickl for reviewing the math.

\bibliographystyle{IEEEtran}

\bibliography{References}

\end{document}